\documentclass[twocolumn,trackchanges]{aastex63}

\usepackage{CJKutf8}
\usepackage{hyperref}

\newcommand{\cntext}[1]{\begin{CJK}{UTF8}{gbsn}#1\end{CJK}\kern-1ex}
\newcommand{\noop}[1]{}

\newcommand{\sdo}{SDO}
\newcommand{\eo}{EOVSA}
\newcommand{\hsi}{RHESSI}
\newcommand{\hinode}{Hinode}

\received{2021 January 20}
\revised{2021 February 8}
\accepted{2021 February 9}
\submitjournal{ApJL}
\turnoffedit

\shorttitle{}
\shortauthors{Chen et al.}
\graphicspath{{./}{figures/}}

\begin{document}

\title{Energetic Electron Distribution of the Coronal Acceleration Region: First results from Joint Microwave and Hard X-ray Imaging Spectroscopy}

\correspondingauthor{Bin Chen}
\email{bin.chen@njit.edu}

\author[0000-0002-0660-3350]{Bin Chen (\cntext{陈彬})}
\affiliation{Center for Solar-Terrestrial Research, New Jersey Institute of Technology, 323 M L King Jr. Blvd., Newark, NJ 07102-1982, USA}

\author[0000-0003-1438-9099]{Marina Battaglia}
\affiliation{University of Applied Sciences and Arts Northwestern Switzerland, 5210 Windisch, Switzerland}

\author[0000-0002-2002-9180]{S\"am Krucker}
\affiliation{University of Applied Sciences and Arts Northwestern Switzerland, 5210 Windisch, Switzerland}

\author[0000-0002-6903-6832]{Katharine K. Reeves}
\affiliation{Harvard-Smithsonian Center for Astrophysics, 60 Garden St., Cambridge, MA 02138, USA}

\author[0000-0001-7092-2703]{Lindsay Glesener}
\affiliation{School of Physics \& Astronomy, University of Minnesota Twin Cities, Minneapolis, MN 55455, USA}

\begin{abstract}
Nonthermal sources located above bright flare arcades, referred to as the ``above-the-loop-top'' sources, have been often suggested as the primary electron acceleration site in major solar flares. The X8.2 limb flare on 2017 September 10 features such an above-the-loop-top source, which was observed in both microwaves and hard X-rays (HXRs) by the Expanded Owens Valley Solar Array (EOVSA) and the Reuven Ramaty High Energy Solar Spectroscopic Imager (RHESSI), respectively. By combining the microwave and HXR imaging spectroscopy observations with multi-filter extreme ultraviolet and soft X-ray imaging data, we derive the energetic electron distribution of this source over a broad energy range from $<$10 keV up to $\sim$MeV during the early impulsive phase of the flare. The best-fit electron distribution consists of a thermal ``core'' from $\sim$25 MK plasma. Meanwhile, a nonthermal power-law ``tail'' joins the thermal core \edit1{at $\sim$16 keV} with a spectral index of $\sim$3.6, which breaks down at above $\sim$160 keV to $>$6.0. In addition, temporally resolved analysis suggests that the electron distribution above the break energy rapidly hardens with the spectral index decreasing from $>$20 to $\sim$6.0 within 20 s, or less than $\sim$10 Alfv\'{e}n crossing times in the source. These results provide strong support for the above-the-loop-top source as the primary site where an on-going bulk acceleration of energetic electrons is taking place very early in the flare energy release. 
\end{abstract}


\section{Introduction} \label{sec:intro}
Hard X-ray (HXR) sources located above bright flare arcades, often referred to as the ``above-the-loop-top'' HXR sources, have been often considered as the primary site for electron acceleration in major solar flares \citep{1994Natur.371..495M, 2008ApJ...673.1181K, 2010ApJ...714.1108K, 2012ApJ...748...33C, 2013ApJ...767..168L, 2014ApJ...780..107K, 2015ApJ...799..129O, 2018ApJ...865...99P}.
These above-the-loop-top HXR sources are mainly due to bremsstrahlung radiation, which sometimes also shows a microwave counterpart \citep{2002ApJ...580L.185M,2010ApJ...714.1108K,2018ApJ...863...83G,2020ApJ...894..158K}. The latter is due to gyrosynchrotron radiation from presumably the same population of accelerated nonthermal electrons gyrating in the coronal magnetic field, an argument corroborated in a recent statistical study by \citet{2020ApJ...894..158K} which found a tight correlation between the $>$50 keV HXR peak flux and the 17 and 34 GHz microwave peak flux from 40 large flares ($>$M7) in solar cycles 23 and 24.

However, previous studies often found that spectral indices of the electron distribution derived from HXR and microwave data are very different, deviating from the simple assumption of an energetic electron distribution with a common power-law shape \citep[e.g.,][]{1994ApJS...90..599K, 2000ApJ...545.1116S, 2013ApJ...763...87A}. It has been argued that such a discrepancy may be attributed to spectral breaks in the electron distribution since HXR and microwave emissions are sensitive to different energy regimes: the typical energy $\varepsilon$ of an electron emitting an HXR photon of energy $\epsilon$ is $\varepsilon\approx1.5\epsilon$--3$\epsilon$, whereas the microwave emission is typically dominated by electrons with energies above $\sim$100 keV \citep{2011SSRv..159..225W,2013ApJ...763...87A}. The discrepancy could also arise from electron transport between spatially separated HXR and microwave sources: the HXRs are usually dominated by precipitated electrons at the footpoints of the flare arcade, while the microwaves are mainly due to trapped electrons within the flare arcade or loop-top region \citep{2008ApJ...673..598M,2012ApJ...747..131K, 2013ApJ...763...87A}.

Important progress has been made by studying partially occulted flares in which the bright footpoint HXR sources are hidden behind the solar limb. These events offer the opportunity to study HXR sources in the corona without the ``contamination'' from the footpoint sources \citep[e.g.,][]{2008ApJ...673.1181K,2010ApJ...714.1108K,2017ApJ...835..124E,2018ApJ...865...99P}. However, the latest microwave spectral imaging observations of the SOL2017-09-10 X8.2-class flare by the Expanded Owens Valley Solar Array (EOVSA; \citealt{2018ApJ...863...83G}) revealed that the microwave emission is present throughout the coronal flaring region: from the loop-top and loop-legs of the flare arcade \citep{2018ApJ...863...83G, 2020ApJ...900...17Y}, to the entire reconnection current sheet and outflow region \citep{2020NatAs...4.1140C}, and even to the footpoints of the erupting flux rope \citep{2020ApJ...895L..50C}. Therefore, imaging spectroscopy in both HXRs and microwaves is required to obtain the spatially resolved spectra from the above-the-loop-top source in isolation. 

Here we report the first spatially resolved HXR and microwave imaging spectroscopy of an above-the-loop-top source, based on the Reuven Ramaty High Energy Solar Spectroscopic Imager (RHESSI; \citealt{2002SoPh..210....3L}) and \eo \ observations of the SOL2017-09-10 X8.2 limb flare during its early impulsive phase. Many aspects of this flare event have already been studied by numerous works. The early impulsive phase is loosely defined as the period around the first HXR and microwave peak at 15:54:20 UT (Figure~\ref{fig:spec}(c) and (f)). It features an erupting cavity and an elongated plasma sheet in the low corona, which is interpreted as, respectively, an erupting magnetic flux rope viewed along its axis and a reconnection current sheet viewed edge on \citep{2018ApJ...853L..18Y,2020ApJ...895L..50C,2020NatAs...4.1140C}. An above-the-loop-top source, observed in both microwaves and HXR, is present above the SXR flare arcade (\citealt{2018ApJ...863...83G}; see also Figure~\ref{fig:spec}(a)). As shown by \citet{2020NatAs...4.1140C}, this source coincides with a local minimum in the magnetic field and the location where most microwave-emitting electrons are concentrated, serving as a ``magnetic bottle'' to confine and accelerate electrons to high energies \citep[see, e.g., recent modeling results in][]{2019ApJ...887L..37K}. Here we combine microwave, EUV, and HXR imaging and spectroscopy to construct the electron spectrum of the above-the-loop-top source over a wide energy range from $<$10 keV up to MeV. We present the imaging spectroscopy results in Section \ref{sec:spec}. We then present the multi-wavelength analyses of the thermal and nonthermal properties of this source in Sections \ref{sec:dem} and \ref{sec:nonthermal}. In Section \ref{sec:discussion}, we discuss their implications in electron acceleration during the early phase of the flare energy release.

\section{Spatially Resolved HXR and Microwave Spectra}\label{sec:spec}
\begin{figure*}[ht!]
\epsscale{1.2}
\plotone{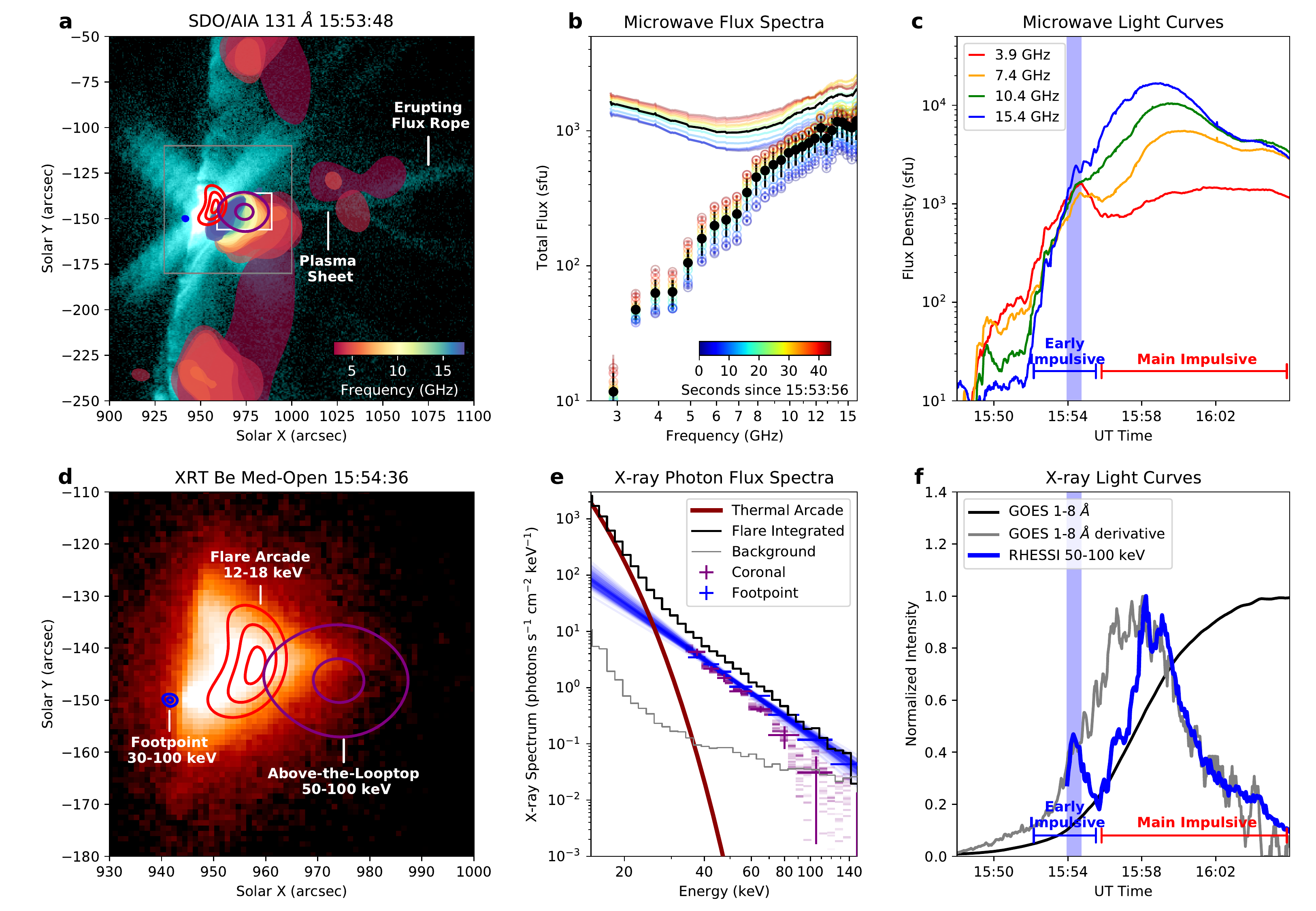}
\caption{Microwave and HXR imaging spectroscopy of the above-the-loop-top source. (a) EOVSA microwave sources (filled contours; 40\% of maximum), RHESSI 12--18 keV flare arcade SXR source (red contours; 30\%, 60\%, 90\% of maximum), RHESSI 30--100 keV footpoint HXR source (blue contours; 50\% and 90\% of maximum), and RHESSI 50--100 keV above-the-loop-top HXR source (purple contours; 50\% and 90\% of maximum) overlaid on \sdo/AIA 131 \AA\ image at 15:53:48 UT. (b) Blue to red solid curves show total-power (full-Sun-integrated) microwave spectra from 15:53:56 UT to 15:54:40 UT. Circle symbols show the microwave spectra from the above-the-loop-top source region (integrated within the white box in (a)). Black symbols indicates the average spectrum within this time interval. (c) Total-power microwave light curves at selected frequencies. (d) More detailed view of the same RHESSI X-ray sources overlaid on \hinode/XRT Be-med image at 15:54:36 UT. The field of view is shown as a gray box in (a). (e) RHESSI X-ray spectroscopy results showing the flare-integrated spectrum (black) and the background spectrum (gray), as well as the spectral component from the thermal arcade (dark red). \edit1{Blue symbols are the spectrum of the footpoint source obtained by imaging spectroscopy. Short purple dashes are the differences between the flare-integrated spectrum and 1,000 trials of power-law fits for the footpoint source (blue lines), representing the emission from the extended coronal source. Purple plus symbols are the average photon flux of the coronal source over each energy band used for imaging spectroscopy.} (f) GOES 1--8 \AA\ SXR light curve (black),  SXR derivative (gray), and RHESSI 50--100 keV HXR light curve (blue). The \edit1{blue strip} in (c) and (f) indicates the time interval shown in (b) during the early impulsive phase. \label{fig:spec}}
\end{figure*}

\hsi\ had four detectors (1, 3, 6, 8) operating during the event. HXR imaging and spectroscopy results during the early impulsive phase of the flare have been already reported in \citet{2018ApJ...863...83G}. Briefly, we select a 48-s time interval 15:53:56--15:54:44 UT (vertical \edit1{blue strip} in Fig.~\ref{fig:spec}(f)) for analysis as it is least impacted by pulse pileup (which occurs when two lower-energy photons arrive at the detector within a short time and are counted by the detector as a single higher-energy photon). X-ray imaging using detectors 1 and 3 in 30--100 keV reveals one compact HXR source located at one footpoint of the bright flare arcade with a full-width-at-half-maximum (FWHM) size of about 3.3$''$. 
This footpoint source coincides with a compact white-light continuum source observed by the Helioseismic and Magnetic Imager aboard the Solar Dynamics Observatory (\sdo/HMI; \citealt{2012SoPh..275..207S}) at this time \citep{2018ApJ...863...83G, 2018ApJ...867..134J}.

A two-step CLEAN procedure \citep{2011ApJ...742...82K} is used to reveal an extended coronal HXR above-the-loop-top source with a FWHM size of $\sim\!20''\times30''$, shown as purple contours in Figures~\ref{fig:spec}(a) and \edit1{(d)}. The coronal source is located above the bright flare arcade seen in both EUV (\sdo/AIA 131 \AA\ shown in Figure \ref{fig:spec}(a), sensitive to $\sim$10 MK plasma) and SXR (\hinode/XRT Be-med image in Figure \ref{fig:spec}(d) and \hsi\ 12--18 keV source shown as red contours).  
Note the coronal source is completely absent (i.e., ``resolved out'') from the image made with detector 3 alone (with a 6.8$''$ resolution), confirming the extended nature of the coronal source\footnote{RHESSI detector modulation is not sensitive to angular scales much larger than the angular resolution of the detector.}. 

HXR imaging spectroscopy is first performed on the footpoint source using detector 3 by imaging at seven HXR energy bands centered at 37.5, 42.5, 47.5, 55, 65,  80, and 105 keV. Figure \ref{fig:spec}(e) shows the best fit of the footpoint photon spectrum (blue crosses) using a power-law function $F(\epsilon)=A_0\epsilon^{-\gamma}$, yielding a photon spectral index of $\gamma=3.4\pm0.2$ and a normalization factor of $\log_{10}(A_0) = 5.9\pm0.3$. The uncertainties are obtained from 1,000 trials of power-law fits after perturbing the data points of the footpoint source by adding random noises within their uncertainties. The differences between the flare-integrated spectrum (black histogram) and the power-law fits of the footpoint spectrum (blue lines) are shown as the short purple dashes. Their average values and standard deviations within each energy band used for footpoint imaging spectroscopy are shown as the purple cross symbols, representing the photon spectrum from the extended coronal HXR source. While the photon spectrum of the coronal source is considerably softer than that of the footpoint source at above $\sim$60 keV, the photon flux of the footpoint and coronal HXR source are comparable at lower energies (within a factor of two at \edit1{$\sim$40--80 keV}). 

\begin{figure*}[ht!]
\epsscale{1.1}
\plotone{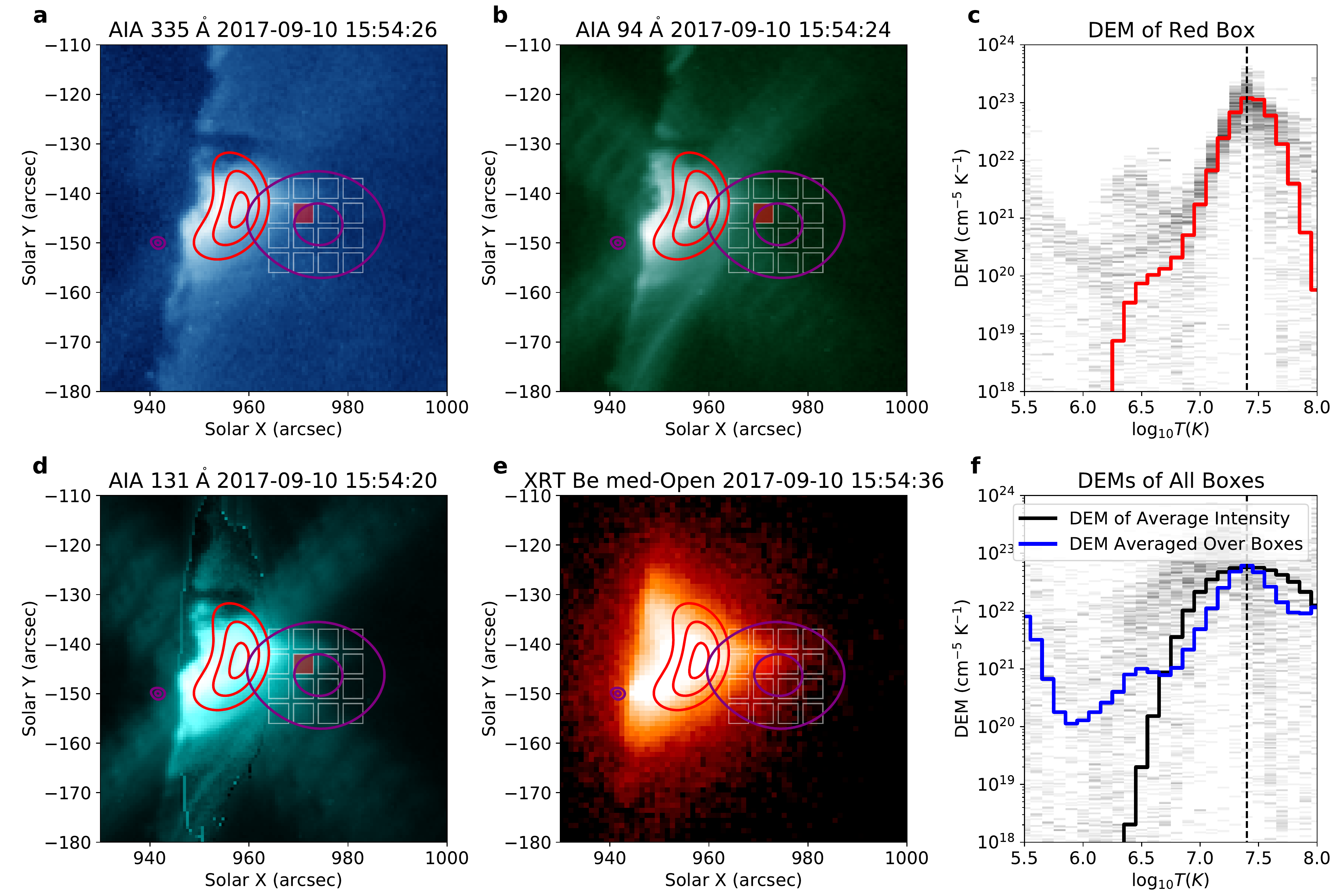}
\caption{Differential emission measure analysis of the above-the-loop-top source by combining \sdo/AIA and \hinode/XRT multi-band images. Panels in the left two columns show four examples of the seven EUV/SXR bands used for the DEM analysis. Contours are the RHESSI X-ray sources same as those in Figure \ref{fig:spec}(d). The sixteen $4''\times 4''$ regions used for deriving the DEM results within the above-the-loop-top region are shown as white boxes. Panel (c) shows an example of the derived DEM curve for a selected region (shown as the red box in left panels). The \edit1{blue curve} in panel \edit1{(f)} shows the \edit1{DEM averaged} over those derived from the 16 small boxes, while the \edit1{black curve} is the DEM derived from the average intensity within the entire above-the-loop-top region. Gray curves in panels (c) and (f) show the Monte Carlo runs. \label{fig:dem}}
\end{figure*}

Microwave imaging spectroscopy is performed at a 4-s cadence using the same method as those described in \citet{2018ApJ...863...83G}. Note the 2.9 GHz data are added in the spectral analysis after adopting the self-calibration technique discussed in \citet{2020ApJ...895L..50C}. In order to perform a joint analysis with the \hsi\ HXR measurements for the same above-the-loop-top source, we integrate the microwave intensity within the same area of the \hsi\ above-the-loop-top source (white box in Figure \ref{fig:spec}(a)) to obtain twelve 4-s-cadence spectra in 15:53:56 UT to 15:54:44 UT (the same time interval used for \hsi\ imaging spectroscopy), shown as color symbols in Figure \ref{fig:spec}(b). The time-averaged microwave spectrum of the above-the-loop-top source within the 48-s RHESSI integration is shown as the black symbols (the length of each error bar corresponds to the standard deviation at the respective microwave frequency). We note a significant difference between the spectra of the above-the-loop-top source and the concurrent total-power (flare-integrated) spectra (solid curves in Figure \ref{fig:spec}(b)) particularly at low frequencies ($<$9 GHz). Such a difference can be understood by recognizing the large spatial extension of the microwave sources to regions beyond the above-the-loop-top area (see \citealt{2020ApJ...895L..50C} for more discussions on the interpretation of the microwave sources in the context of the standard flare scenario depicted in three dimensions). It vividly demonstrates the limitations of microwave spectral analysis using the total-power spectrum alone.

\section{Differential Emission Measure Analysis}\label{sec:dem}
In order to constrain the thermal properties of the above-the-loop-top source region, we perform differential emission measure (DEM) analysis by combining images from six \sdo/AIA passbands (94, 131, 171, 193, 211, and 335 \AA) at around 15:54:20 UT and the \hinode/XRT Be-med image obtained at the closest time (15:54:36 UT). These EUV and SXR bands combined provide a broad temperature sensitivity of coronal plasma from $<$1 MK up to $\lesssim$100 MK \citep{2007SoPh..243...63G,2012SoPh..275...17L}. The DEM analysis was carried out using a robustly tested DEM reconstruction routine {\tt xrt\_dem\_iterative2} \citep{2004ASPC..325..217G,2004IAUS..223..321W}, which returns a distribution of the DEM as a function of temperature $T$, i.e., $\xi(T) = d(n_{\rm th}^2 d)/dT$ (where $n_{\rm th}$ is the thermal electron density and $d$ is the column depth) from selected spatial regions in the image. Coronal abundances are assumed for the above-the-loop-top source, following the analysis made by \citet{2018ApJ...854..122W} based on data from SDO/AIA and the EUV Imaging Spectrometer (EIS) aboard Hinode.

We divide the above-the-loop-top region into sixteen $4''\times4''$ small boxes (Figure \ref{fig:dem}) and take the average intensities within each small box to perform the DEM analysis. A representative DEM distribution of a small box located near the centroid of the above-the-loop-top HXR source is shown in Figure \ref{fig:dem}(c), which displays a prominent peak at $T\approx25$ MK ($\log_{10}T=7.4$). Monte Carlo runs (gray curves in Figure \ref{fig:dem}(c)) confirm that this high-temperature peak in the resulting DEM curve is robust. To further verify this high-temperature peak, we also performed DEM analysis using the average intensities of the entire above-the-loop-top region, which returns the same peak at $\sim$25 MK albeit with larger uncertainties owing to the more substantial intensity variations across the region. The average column emission measure is $\sim\!2\times10^{30}$ cm$^{-5}$. Taking a column depth $d$ with the same value as the size of the above-the-loop-top source (20$''$--30$''$), the average thermal electron density of the above-the-loop-top source is about 3--4$\times 10^{10}$ cm$^{-3}$. Our DEM analysis results are broadly consistent with those derived from EUV spectroscopy data made by the Extreme-Ultraviolet Imaging Spectrometer aboard \hinode\  \citep{2018ApJ...854..122W, 2018ApJ...864...63P}, although the latter used measurements made during the main peak of the flare (5--10 minutes after our time of interest).

\section{Joint HXR and Microwave Spectral Analysis}\label{sec:nonthermal}
The HXR and microwave spectra of the above-the-loop-top source obtained from spatially resolved \hsi\ and \eo\ data offer a unique opportunity of reconstructing the underlying energetic electron distribution over a wide energy range. As introduced in Section \ref{sec:intro}, the HXR spectrum provides excellent constraints for the nonthermal electrons at relatively \edit1{low} energies ($\lesssim$100 keV in this event), while the microwave data complement the HXR diagnostics with added constraints for the more energetic, $\gtrsim$100 keV electrons. In addition, the optically-thick part of the microwave spectrum also provides diagnostics for the thermal electron density, complementing the DEM analysis discussed in Section \ref{sec:dem}.

For the nonthermal electron density distribution $f(\varepsilon)=dn_{\rm nth}(\varepsilon)/d\varepsilon$, we adopt a broken power-law functional form with free parameters including the total electron density $n_{\rm nth}$, low-energy cutoff $\varepsilon_{\rm min}$, break energy $\varepsilon_{\rm b}$, and spectral indices $\delta'_{1}$ and $\delta'_{2}$ below and above the break energy, respectively. Additional parameters used for the joint spectral analysis include the magnetic field strength $B$, thermal electron density $n_{\rm th}$, and viewing angle $\theta$ (with regard to the magnetic field direction). The thermal temperature $T$ of the above-the-loop-top source is fixed to 25 MK according the DEM results and the column depth $d$ is set to $20''$.

To forward-fit the HXR emission, we assume the thin-target bremsstrahlung model and use the Python codes available in the SunPy package {\tt sunxspex}\footnote{\url{https://github.com/sunpy/sunxspex/blob/master/sunxspex/emission.py}, which is a Python version of the standard routine {\tt brm2\_thintarget.pro} available in SolarSoft IDL.}. For the microwave emission, we forward-calculate the gyrosynchrotron radiation from the broken power-law electron distribution by adopting the fast gyrosynchrotron codes developed by \citet{2010ApJ...721.1127F}. We use a global minimization method \texttt{differential evolution} available in Scipy's {\tt optimize} package to fit the observed HXR and microwave spectra jointly. We note that the number of the independent measurements used for the minimization (35) is much greater than the degrees of freedom (8) adopted in the joint fit, therefore constituting a well-determined minimization problem. 

\begin{figure*}[ht!]
\epsscale{1.1}
\plotone{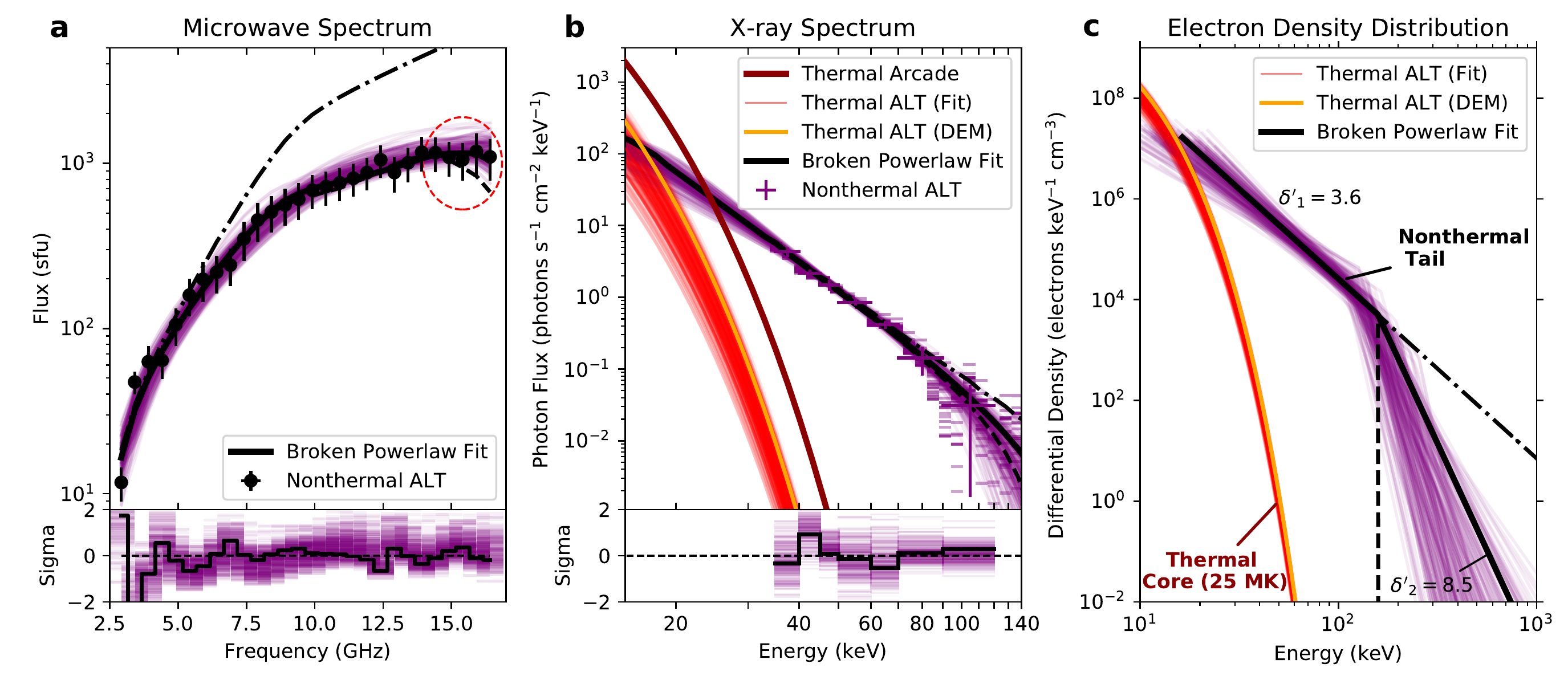}
\caption{Joint HXR and microwave spectral fit of the above-the-loop-top source (``ALT'' in the figure legends). (a) Best fit for the microwave spectrum (thick black curve) and the residuals. Microwave spectra from 200 of 1000 MCMC samples of the parameter space and the corresponding residuals are shown as thin purple curves. (b) Best fit for the above-the-loop-top X-ray source (thick black curve) and the residuals. Thin purple curves are the calculated HXR spectra from the same MCMC samples. Thin-target bremsstrahlung is assumed for the nonthermal HXR emission. The corresponding thermal bremsstrahlung X-ray spectra from the MCMC samples and DEM analysis are shown as the thin red curves and thick orange curve, respectively. (c) Corresponding best-fit and MCMC samples of the thermal component (red curves) and broken powerlaw electron density distribution (thick black curve and thin purple curves). The DEM-derived thermal electron density distribution is shown as the thick orange curve. Also shown are the calculated microwave and HXR spectra for two test cases of the nonthermal electron distribution: dash-dot curves are for a power-law spectrum that does not break at $\varepsilon_b$, and dashed curves are for a spectrum that cuts off completely at $\varepsilon_b$. See text for discussions. \label{fig:spec_fit}}
\end{figure*}

\begin{figure*}[ht!]
\epsscale{1.1}
\plotone{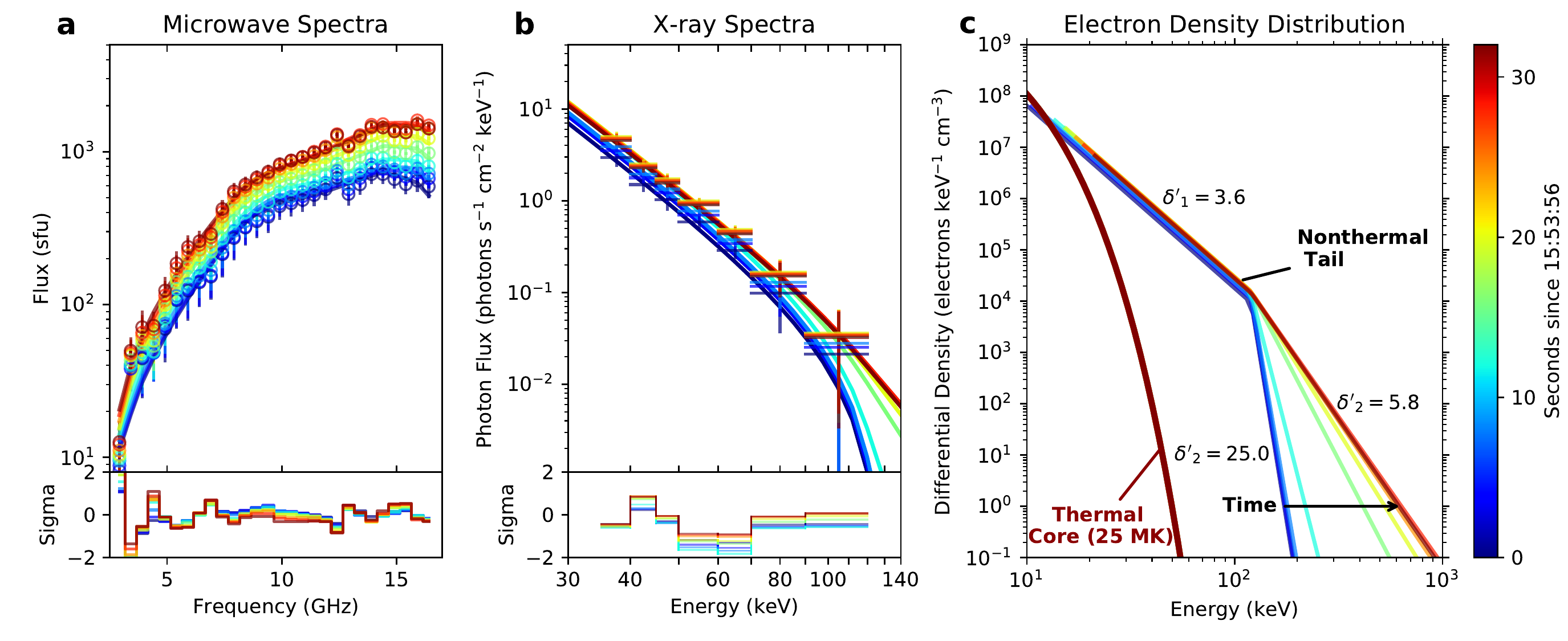}
\caption{Temporal evolution of the above-the-loop-top source and the corresponding fit results. (a) Temporally resolved EOVSA microwave spectra of the above-the-loop-top source from 15:53:56 UT to 15:54:28 UT when an increasing microwave flux is present at all frequencies (circle symbols colored in time from blue to red) and their best fits (solid curves). (b) RHESSI HXR spectra based on the 48-s averaged spectrum of the above-the-loop-top source but scaled in time according to the flare-integrated 50--100 keV light curve. Best-fit results are shown as solid curves. (c) The corresponding time evolution of the best-fit nonthermal electron distribution shown in the same color scheme. \label{fig:spec_t2t}}
\end{figure*}

To test the robustness of the fit results, we also employ a Bayesian-statistics-based Markov chain Monte Carlo (MCMC) method (implemented in an open-source software {\tt emcee} included in the Python package {\tt lmfit}\footnote{\url{https://lmfit.github.io/lmfit-py}}), to evaluate the probability distributions of the fit parameters, which are in turn used to estimate the associated uncertainties (Figure~\ref{fig:mcmc}; see Appendix for details). 
The best-fit results for both the microwave and HXR spectra, shown as thick black curves in Figure \ref{fig:spec_fit}(a) and (b), respectively, agree very well with the distributions of the MCMC runs (thin purple curves). The corresponding best-fit and MCMC runs of the broken power-law electron distribution are shown in Figure \ref{fig:spec_fit}(c). The thermal electron spectra returned from the MCMC runs of the joint fit are also displayed as thin red curves, in agreement with the DEM analysis results (thick orange curve). 

As expected, the electron spectral index below the break energy $\delta'_1=3.6^{+0.1}_{-0.7}$ is quite nicely constrained, thanks to the HXR measurements with relatively small uncertainties below $\sim$100 keV. The break energy $\varepsilon_b$ is at 158$^{+14}_{-46}$ keV, above which the spectral index of the higher energy electrons shows a significant break-down, with $\delta'_2 > 6.0$ (the lower-bound of the probability distribution; see Figure~\ref{fig:mcmc}). The best-fit thermal electron density is $n_{\rm th}\approx 2.4^{+0.9}_{-0.6}\times10^{10}$ cm$^{-3}$ (red curves in Figure~\ref{fig:spec_fit}(c) show the corresponding electron distribution), consistent with that estimated from the DEM analysis ($n_{\rm th} \approx 3.8 \times 10^{10}$ cm$^{-3}$ with a column depth of $d = 20''$; orange curve in Figure~\ref{fig:spec_fit}(c)). The X-ray emission expected from the thermal component of the above-the-loop-top source, shown in Figure~\ref{fig:spec_fit}(b) as red and orange curves (calculated using the microwave- and DEM-constrained thermal electron density, respectively, and $T=25$ MK), is ``buried'' under the X-ray emission from the thermal flare arcade (dark red curve in Figure~\ref{fig:spec_fit}(b)) in the flare-integrated RHESSI X-ray spectrum. Such a weak thermal emission also renders it difficult to detect in RHESSI SXR images with the presence of the bright flare arcade (c.f., the 12--18 keV image shown in Figure~\ref{fig:spec}(d)). In the Hinode/XRT Be-med image (which is sensitive to a broad temperature range of $\sim$10--100 MK; \citealt{2007SoPh..243...63G}), this 25 MK above-the-loop-top source may correspond to the faint tip of a cusp-shaped structure above the bright SXR flare arcade. 

Remarkably, in the electron density distribution of the above-the-loop-top source, this 25 MK thermal ``core'' is joined by the nonthermal electron spectrum at $\sim$16 keV nearly seamlessly, implying a continuous electron population resulted from the bulk energization process. \edit1{We note that while it is rather challenging to constrain the low-energy cutoff $\varepsilon_{\rm min}$ of the nonthermal electron spectrum using the HXR data alone (\citealt{2011SSRv..159..107H}; see also recent developments in \citealt{2019ApJ...871..225K} and references therein with the warm-target approach), the microwave spectrum is sensitive to, among others, the total nonthermal electron density $n_{\rm nth}$ \citep[see, e.g., Movie S2 in][]{2020Sci...367..278F}. The latter depends strongly on $\varepsilon_{\rm min}$, particularly when a tight constraint for $\delta'_1$ and the nonthermal electron density above $\sim$40 keV is already available from the HXR diagnostics. Therefore, our joint microwave--HXR fit complements each other to provide an optimal constraint for both the low-energy cutoff $\varepsilon_{\rm min} = 16^{+3}_{-6}$ keV and the total nonthermal electron density $n_{\rm nth}=1.1^{+0.5}_{-0.4}\times 10^8\ \mathrm{cm}^{-3}$}. The magnetic field strength returned from the fit is 845$^{+198}_{-47}$ G, \edit1{consistent with earlier findings of a strong coronal magnetic field in this flare (and the associated active region) by using direct spectroscopy measurements \citep{2018ApJ...863...83G,2019ApJ...874..126K,2020Sci...367..278F,2020NatAs...4.1140C} and indirect extrapolation estimates \citep{2018ApJ...868..148L,2019ApJ...880L..29A}.} The corresponding plasma $\beta=P_{\rm gas}/P_{\rm B}=8\pi n_{\rm th}kT/B^2$ is $\sim$0.003 and the Alfv\'{e}n speed $v_A=B/\sqrt{4\pi n_{\rm th}\mu m_H}$ is around 10,000 km s$^{-1}$. 

The steep break-down of the electron spectrum above $\sim$160 keV resulting from the joint HXR/microwave fit is mainly determined by the observed microwave spectrum. To demonstrate the necessity of such a spectral break down, we calculate the HXR and microwave spectra from two limiting cases of the electron distribution. Case A: A single power-law having the same best-fit spectral index of $\delta'_1$ but extending beyond $\varepsilon_b$ to large energies (dash-dot curve in Figure~\ref{fig:spec_fit}(c)); Case B: The same power-law spectrum that cuts off completely at $\varepsilon_b$ (dashed curve in Figure~\ref{fig:spec_fit}(c)). Case A results in a microwave spectrum that greatly exceeds the observed flux at above $\sim$5 GHz due to the excessive high-energy electrons (dash-dot curve in Figure~\ref{fig:spec_fit}(a)). Case B produces a microwave spectrum that is very close to the best-fit case at most frequencies except the few highest frequency data points (dashed curve in Figure~\ref{fig:spec_fit}(a); the deviation is marked by an oval in Figure~\ref{fig:spec_fit}(a)). The latter is due to the unavailability of the electrons with sufficiently high energy to boost the microwave emission toward these high frequencies. The limited number of data points available to constrain the lower limit of the high-energy electron population above $\varepsilon_b$ also explains the relatively loose constraints on the upper bound of $\delta'_2$, as demonstrated by the MCMC results. In either case, the corresponding HXR spectrum shows minor differences at the highest measured energies that are nearly indistinguishable within the uncertainties (dash-dot and dashed curves in Figure~\ref{fig:spec_fit}(b)).

\section{Discussions} \label{sec:discussion}
In the previous sections, we have combined microwave, EUV, and X-ray imaging and spectroscopy data of the same coronal above-the-loop-top source observed during the early impulsive phase of the SOL2017-09-10 flare to derive a comprehensive, and nearly continuous energetic electron spectrum from $<$10 keV to $\sim$1 MeV. The best-fit spectrum shows a steep break down at above $\sim$160 keV from $\delta'_1\approx3.6$ to $\delta'_2>6.0$. Such a spectral break down is an indication for the relatively small number of microwave-emitting, mildly relativistic electrons very early in the flare energy release. 

However, several minutes later when the microwave/HXR fluxes reach their peak at $\sim$15:58--15:59 UT (c.f., Figures \ref{fig:spec}(c) and (f)), the spectral index of the above-the-loop-top region derived from the microwave spectra hardens significantly to $\delta'\approx 3.6$ (Fleishman et al., in preparation). The flare-integrated HXR spectrum also displays a broken powerlaw that appears to break up at above $\sim$50 keV with a photon spectral index of $\gamma\approx 3.3$ \citep{2019RAA....19..173N}, which corresponds to $\delta'\approx 2.8$ of the electron density distribution for thin-target bremsstrahlung and $\delta'\approx 4.8$ for thick-target bremsstrahlung, although the significant pileup effects during the later times complicate the quantitative analysis and interpretation. These observations imply an ongoing acceleration process that quickly energizes $\gtrsim$100 keV electrons and hardens the electron distribution within a few minutes.

It is beyond the scope of the current work to expand our joint microwave/HXR spectral analysis to these later times, particularly because a significant pileup effect exists for the HXR data. However, the temporally resolved microwave imaging spectroscopy data does allow us to gain some insights on the temporal evolution of the high-energy electrons within this short interval during the early impulsive phase. Since RHESSI imaging spectroscopy does not provide temporally resolved HXR spectra for the above-the-loop-top source during this period, as a first-order approximation, we scale the absolute HXR photon flux of the 48-s-averaged spectrum (i.e., purple crosses in Figure~\ref{fig:spec}(e)) as a function of time according to the temporal variation of the flare-integrated 50--100 keV count rate in Figure~\ref{fig:spec}(f), producing the pseudo temporally resolved HXR spectra shown in Figure~\ref{fig:spec_t2t}(b). For simplicity, we further fix the \edit1{density of the thermal electrons $n_{\rm th}$}, viewing angle $\theta$, and the spectral index below the break $\delta'_1$ to the best-fit values from the time-averaged spectra, and set the break energy to $\varepsilon_b = 120$ keV. Figure \ref{fig:spec_t2t} shows the time-dependent fit results and the corresponding electron density distribution, \edit1{colored} in time from blue to red for the interval from 15:53:56 to 15:54:30 UT when the microwave fluxes at all frequencies are growing.  
In accordance with the increasing microwave flux, the high-energy electron population above $\varepsilon_b$ experiences a rapid hardening: the electron spectral index above the break $\delta'_2$ decreases from $>$20 to $\sim$6 within 20 s, or $\lesssim$10 Alfv\'{e}n crossing times $\tau_A$ in the above-the-loop-top source region ($\tau_A = L/v_A$, where $L=20''$--$30''$ is the source size). Such a rapid spectral hardening of nonthermal electrons within several Alfv\'{e}n crossing times has been found in recent particle acceleration simulations for macroscale low plasma $\beta$ systems \citep{2016ApJ...820...60G,2018ApJ...866....4L,2020arXiv201101147A}. We caution that, however, the results demonstrated here are subject to the validity of the assumptions we adopt: both the fraction of the HXR flux in the above-the-loop-top source and the \edit1{spectral shape of the HXR photon spectrum (which is mainly determined by $\delta'_1$)} remain unchanged during this 34-s period, which cannot be examined in detail due to the unavailability of HXR imaging spectroscopy at a finer time resolution.

In our analysis, the thin-target bremsstrahlung scenario is assumed. The assumption is largely valid since the critical energy for stopping the electrons in the above-the-loop-top source due to Coulomb collisions $\varepsilon_{c}\approx8.8N_{19}^{1/2}\lesssim 20$ keV (where $N_{19}=n_{\rm th}L$ is the column density in units of $10^{19}$ cm$^{-2}$), well below most energies in the nonthermal electron spectrum. However, under the strong diffusion limit for which electrons undergo a random-walk-type transport process, the HXR above-the-loop-top source can become a thick target \citep{1999ApJ...522.1108M,2013A&A...551A.135S,2018ApJ...865...99P}. Nevertheless, such a coronal thick-target scenario is deemed unlikely for our case because of the presence of a bright 30--100 keV footpoint source, which indicates abundant precipitated electrons at $>$30 keV. An intermediate thin-thick target scenario may still be possible in the case of a transitional weak-to-strong diffusion. A detailed investigation is beyond the scope of this Letter. 

Although a broken power-law function for the nonthermal electron distribution is found to agree very well with the observed microwave and HXR spectra, our results do not necessarily exclude the possibility of a different type of source electron distribution. Distributions that are intense and flat at low energies, but weaker and steeper at high energies, may also work. One such example is the kappa distribution \citep{2015ApJ...799..129O,2015ApJ...815...73B,2019ApJ...872..204B}. In addition, despite that our data provide direct diagnostics for the above-the-loop-top source in isolation, transport effects within the above-the-loop-top source (e.g., trapping, scattering, collisional loss) may alter the ``pristine'' flare-accelerated electron spectrum, leading to a spectral break-down at higher energies \citep{1976MNRAS.176...15M, 1995SoPh..158..283W, 1998ApJ...505..418F,1999ApJ...522.1108M,2018ApJ...865...99P}. While care must be taken when interpreting the observations, the rapid hardening of the electron spectrum does suggest a likely ongoing-acceleration. Last but not least, in our analysis, an isotropic electron angular distribution is assumed. Amendments need to be made if a significant anisotropy of the electron distribution is present \citep{2003ApJ...587..823F,2004ApJ...613.1233M,2007A&A...466..705K,2012ApJ...750...35C}\edit1{, although observational evidence for the anisotropy in the above-the-looptop sources has been elusive}. Our data do not provide adequate constraints for distinguishing these above scenarios. Further progress calls for next-generation microwave and HXR instrumentation that can provide high dynamic range imaging spectroscopy observations along with simultaneously high spatial, temporal, and spectral resolution, such as the Frequency Agile Solar Radiotelescope (FASR; \citealt{2019astro2020U..56B}) and a spacecraft version of the Focusing Optics X-ray Solar Imager sounding rocket (FOXSI; \citealt{2013SPIE.8862E..0RK}).

\newpage

\acknowledgments

We are grateful to Lyndsay Fletcher, Eduard Kontar, Dale Gary, Gregory Fleishman, James Drake, Harry Arnold, Fan Guo, and Xiaocan Li for helpful discussions within the SolFER DRIVE Science Center collaboration. The work is supported partly by NASA DRIVE Science Center grant 80NSSC20K0627. EOVSA operation is supported by NSF grant AST-1910354. B.C. is supported by NSF grant AGS-1654382 and NASA grant 80NSSC20K1318 to NJIT.
K.R. is supported by NSF grant AGS-1923365 to SAO. 
L.G. is supported by NASA grant 80NSSC20K1277.
Hinode is a Japanese mission developed and launched by ISAS/JAXA, with NAOJ as domestic partner and NASA and STFC (UK) as international partners. It is operated by these agencies in cooperation with ESA and NSC (Norway).

\facilities{OVRO:SA, RHESSI, SDO, Hinode}

\software{Astropy \citep{2018AJ....156..123A},
          CASA \citep{2007ASPC..376..127M},
          LMFIT \citep{2014zndo.....11813N},
          NumPy \citep{harris2020array},
          SciPy \citep{2020NatMe..17..261V},
          SunPy \citep{2020ApJ...890...68S},
          }

\newpage

\appendix
\label{ap}
\section{Markov Chain Monte Carlo Analysis of the Joint Spectral Fit}
\begin{figure*}[ht!]
\epsscale{1.2}
\plotone{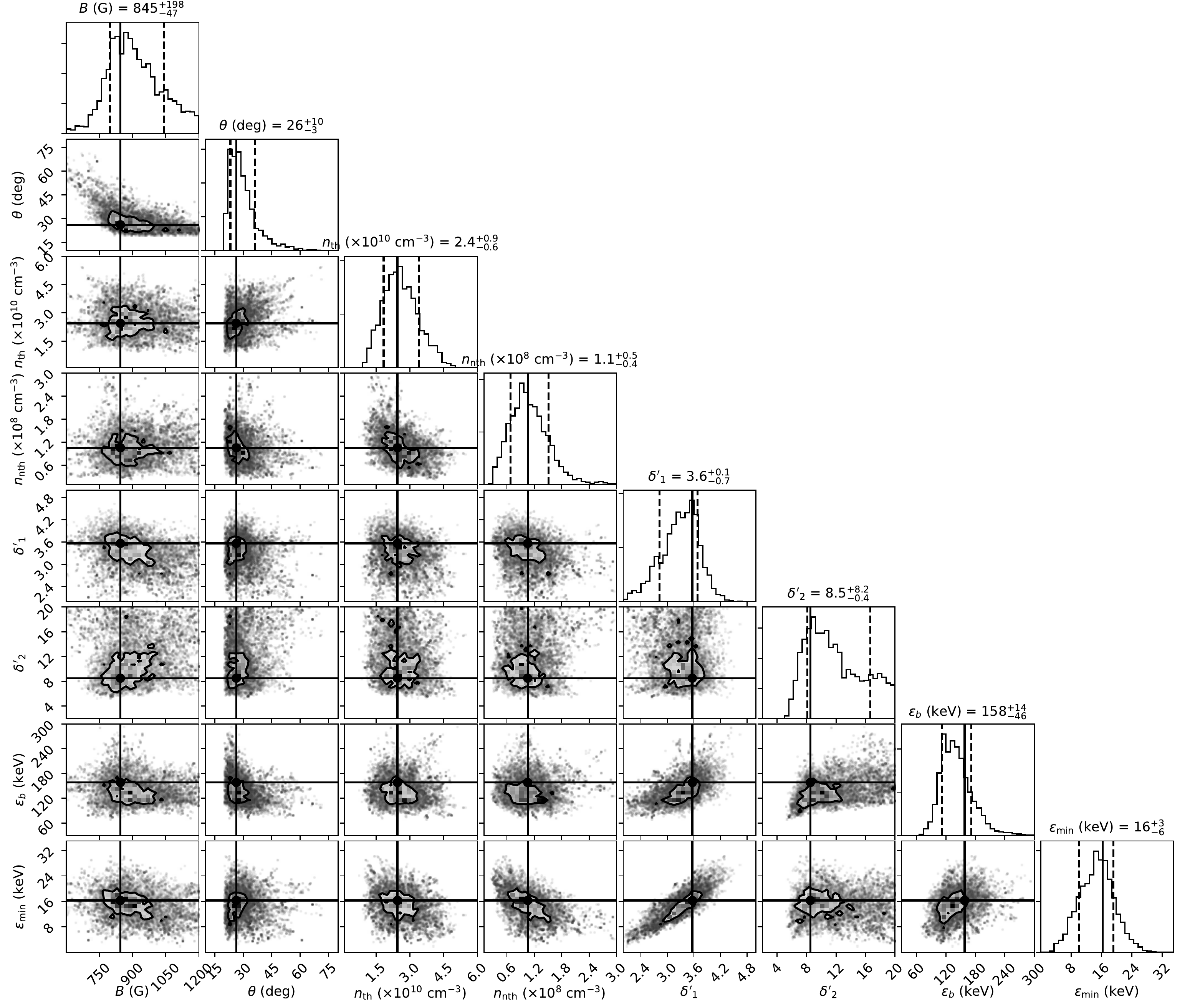}
\caption{Markov chain Monte Carlo analysis of the joint microwave and HXR fit results. See Appendix for details. \label{fig:mcmc}}
\end{figure*}

We employ a Markov chain Monte Carlo (MCMC) method to evaluate the fit results. The technique and software used for the MCMC analysis are identical to those described in the Methods section of \citet{2020NatAs...4.1140C}. 
Figure~\ref{fig:mcmc} shows the MCMC analysis results in the form of a corner plot, in which the diagonal panels show the one-dimensional projections of the probability distributions of the respective fit parameters. The two-dimensional histograms of the probability distributions between pairs of the fit parameters are shown as the non-diagonal panels. \edit1{Solid horizontal/vertical lines in each panel indicate the best-fit values from the minimization.} The widths of the distributions provide optimal estimates for the uncertainties of the respective fit parameters. The uncertainties shown for each fit parameter are estimated using the 16\% and 84\% quantiles of the respective 1-D histograms, which correspond to approximately the 1-$\sigma$ level ($1-e^{-0.5} \approx 39.3\%$) in the corresponding 2-D histograms (solid contours). 

\bibliography{looptop20}{}

\end{document}